\documentclass{rmf-d}
\usepackage{nopageno,rmfbib,multicol,times,epsf,amsmath,amssymb,cite}
\usepackage[latin1]{inputenc}
\usepackage[]{caption2}
\usepackage[]{graphicx}
\usepackage{lipsum} 
\usepackage{nonfloat}
\newcommand\myfigure[1]{%
\medskip\noindent\begin{minipage}{\columnwidth}
\centering%
#1%
\end{minipage}\medskip}
\def\rmfcornisa{Research OR Education of Physics \hfill\rmf\ {\bf ??} (*?*) ???--???
\hfill MES? A\~NO?}

\def\rmfcintilla{{\it Rev.\ Mex.\ Fis.\/} {\bf ??} (*?*) (????) ???--???}
\clearpage \rmfcaptionstyle 
\begin{document}
\title{
Parton distribution effects in the direct photon production at hadron collisions
\vspace{-6pt}}
\author{David F. Renter\'ia-Estrada\footnote{Speaker},}
\address{Facultad  de  Ciencias  F\'isico-Matem\'aticas,  Universidad  Aut\'onoma  de  Sinaloa,  Ciudad  Universitaria, CP 80000 Culiac\'an, M\'exico.}
\author{Roger J. Hern\'andez-Pinto}
\address{Facultad  de  Ciencias  F\'isico-Matem\'aticas,  Universidad  Aut\'onoma  de  Sinaloa,  Ciudad  Universitaria, CP 80000 Culiac\'an, M\'exico.}
\author{German F. R. Sborlini}
\address{Deutsches Elektronen-Synchrotron DESY, Platanenallee 6, 15738 Zeuthen, Germany.}

%\author{Email: davidrenteria.fcfm@uas.edu.mx, roger@uas.edu.mx and german.sborlini@desy.de}
\maketitle
\recibido{day month year}{day month year
\vspace{-12pt}}
\begin{abstract}
\vspace{1em}
Parton distribution functions are crucial to understand the internal kinematics of hadrons. There are currently a large number of distribution functions on the market, and thanks to today's technology, performing computational analysis of the differential cross-sections has become more accessible. Despite technological advances, accurately accessing to the internal structure of hadrons remains a difficult task from a theoretical point of view. In this work, we analyze the impact on the differential cross-sections when updating the sets of parton distribution and fragmentation functions, for the production of one hadron plus a direct photon at the energy scale of RHIC and LHC experiments.
\vspace{1em}
\end{abstract}
\keys{NLO Computation, QCD Phenomenology, High Energy Physics  \vspace{-4pt}}
\pacs{   \bf{\textit{14.70.Bh, 14.40.Aq, 02.70.Ss, 13.90.+i}}    \vspace{-4pt}}
\begin{multicols}{2}

%%%%%%%%%%%%%%%%%%%%%%%%%%%%%%%%%%%%%%%%%%%%%%%%%%%%%%%%%%%%%%%%%%%%%%%%%%%%%%%%%%%%%%%%%%%%%%%%%%%%%%%%
\section{Introduction}
Nowadays, there are different methods to perform the reconstruction of the parton kinematics in hadronic collisions, which is important to understand their internal structure. In addition, it is necessary to deal with complex mathematical models whose solutions are difficult to obtain analytically. So, it is recurrent to use perturbative methods to find approximated solutions. One way to access the internal structure of particles at hadron collision, is through the study of the distribution of quarks and gluons inside the original hadrons. These distributions are called parton distribution functions (PDF) and are extracted from the experiments by using advanced fitting and modelling methods \cite{AbdulKhalek:2019bux}. It is possible to define such distributions for both unpolarized (i.e. spin-averaged) and polarized observables.

It is well known that deep inelastic scattering (DIS) experiments are not enough to constrain the shape of polarized quark and gluon distributions \cite{deFlorian:2008mr,deFlorian:2009vb}. Nowadays, the \emph{proton spin crisis} is an active topic and the scientific community still did not find the solution. Therefore having a precision description of these parton distributions, is crucial to address this problem \cite{Cruz-Martinez:2021rgy}. In order to find possible solutions to the spin crisis and obtain more information about the internal dynamics of hadrons, we need to access to the parton level kinematics in the most clean and unperturbed way.

The main problem to access the kinematics of the partons is presence of a dense and hot medium caused by QCD particle interactions in high-energy collisions. Moreover, such states cannot be easily described within the perturbative approach. On the other hand, thanks to the low interaction that photons have with the medium, the measurement of hard photons in the final state becomes a good alternative for the solution to this problem.  

In this article, we center our attention on the impact of the PDFs in the production of a direct photon plus one hadron including up to NLO QCD correction. This work is based on studies carried out in Refs. \cite{deFlorian2011PRD83,RenteriaEstrada2021sym13,RenteriaEstrada2021arxiv}, where the last two articles constitutes a step toward a more complete and updated description of the phenomenology of the production hadron plus photon at colliders. More recent studies \cite{RenteriaEstrada2021arxiv} make use the latest PDFs to reconstruct the internal kinematics of the partons using neural network methods in machine learning. Furthermore, hard photons were considered in other works in order to establish patterns of energy loss in heavy ion collisions \cite{Wang:1996yh}, analyze the sensitivity to medium-induced modifications in the Fragmentation Functions (FF) \cite{Arleo:2004xj, Arleo:2006xb, Zhang:2009rn}, and other studies. Thus, in Sec. \ref{sec:computation}, we explain the theoretical framework applied to calculate NLO QCD correction to the differential cross-section for the process $pp \to h + \gamma$. The methodology that we have implemented to distinguish hard photons is based on the smooth isolation algorithm, which allows to consistently suppress events with photons originated in electromagnetic decays of hadrons. In Sec. \ref{sec:results}, we explain the cuts implemented for relevant experiments at RHIC and LHC energies, and we compare our results using new PDFs sets with respect to the study performed in Ref. \cite{deFlorian2011PRD83}. Finally, in Sec. \ref{sec:conclusions}, we summarize this presentation and explain future strategies to explore the inner structure of hadrons by using this observable. 

%20211229: REV GS

%%%%%%%%%%%%%%%%%%%%%%%%%%%%%%%%%%%%%%%%%%%%%%%%%%%%%%%%%%%%%%%%%%%%%%%%%%%%%%%%%%%%%%%%%%%%%%%%%%%%%%%%
\section{Computation of cross-section }
\label{sec:computation}
The cross-section calculation relies on the parton model, which allows to describe the collisions between hadrons in the high-energy regime in terms of fundamental particles. Due to the factorization theorem \cite{Collins:1989gx}, in this kinematic regime, perturbation theory can be applied. Explicitly, the cross-section is described by the convolution of the PDF, the FF, and the partonic cross-section. Therefore, in the case of producing a direct photon plus one specific hadron, the cross-section is described by,
\begin{eqnarray}
\nonumber d\sigma_{H_1 \, H_2 \to h \, \gamma}^{\rm DIR} &=& \sum_{a_1 a_2 a_3} \int dx_1 dx_2 dz \, f^{(H_1)}_{a_1}(x_1,\mu_I)\nonumber \\
&\times\,& f^{(H_2)}_{a_2}(x_2,\mu_I)\,d^{(h)}_{a_3}(z,\mu_F) d\hat\sigma^{\rm DIR}_{a_1\,a_2 \to a_3 \, \gamma} \, ,\nonumber \\ 
\label{eq:Direct}
\end{eqnarray}
with $H_1$ and $H_2$ the hadrons colliding in the initial state, $a_i$ the partons involved in the process and $d\hat\sigma^{\rm DIR}$ the differential partonic cross-section. The function $f^{(H)}_{a}(x,\mu_I)$ represents the PDF associated to the collinear emission of a parton of flavor $a$ from the hadron $H$ with momentum fraction $x$ at the initial factorisation scale $\mu_I$. Analogously, $d^{(h)}_{a}(z,\mu_F)$ represents the density probability function of generating a hadron $h$ with momentum fraction $z$ from the parton $a$, at the final factorisation scale $\mu_F$. Regarding the scale dependence, the partonic cross-section includes terms depending on $\mu_I$, $\mu_F$ and also on the renormalization scale, $\mu_R$.
Eq. (\ref{eq:Direct}) describes a measure of the probability of producing \emph{prompt} photons, i.e. photons that are emitted directly by the interaction of the originally colliding partons. However, in real life,  detectors cannot distinguish the nature of the photon. Therefore, it is be necessary to consider the contribution of detecting a photon produced by the decay of another particle. To do this, we must also compute the fragmentation or \emph{resolved} component of the cross-section, i.e.
\begin{eqnarray}
\nonumber d\sigma_{H_1 \, H_2 \to h \, \gamma}^{\rm RES} &=& \sum_{a_1 a_2 a_3 a_4} \int dx_1 dx_2 dz dz' \, f^{(H_1)}_{a_1}(x_1,\mu_I)
\\ &\times &f^{(H_2)}_{a_2}(x_2,\mu_I)\, d^{(h)}_{a_3}(z,\mu_F) d^{(\gamma)}_{a_4}(z',\mu_F)
\nonumber\\ &\times& d\hat\sigma_{a_1\,a_2 \to a_3 \, a_4}  \, ,
\label{eq:Resolved}
\end{eqnarray}
where the parton $a_4$ generates a photon after hadronization. Notice the presence of the parton-to-photon fragmentation function, $d^{(\gamma)}_{a}(z,\mu_F)$. This quantity is not very well constrained experimentally, due to non-perturbative and low-energy effects. This separation of components is not a physical result. However, the analysis to be performed requires having a predominant contribution due to direct photons in the final state. The way we attack this problem, is suppressing by the resolved component by means of so-called isolation prescriptions. 

%20211229: REV GS

%%%%%%%%%%%%%%%%%%%%%%%%%%%%%%%%%%%%%%%%%%%%%%%%%%%%%%%%%%%%%%%%%%%%%%%%%%%%%%%%%%%%%%%%%%%%%%%%%%%%%%%%
\subsection{Implementation of the isolation criteria}
\label{ssec:Isolation}
In general, the interactions between QCD partons generate a dense medium of highly energetic particles, which are rapidly hadronized. When a photon is emitted as a consequence of the decay of hadrons or strongly-interacting particles, it is expected to be accompanied by a bunch of hadrons. On the other hand, when the photon is generated directly from the partonic collision, it should produce a clear and isolated signal in the detector. By definition, isolated photons are those that fulfill certain selection criteria, establishing a separation from this particle to any hadron or jet. Usually, distances are measured within the rapidity-azimuthal plane: if $a=(\eta_1,\phi_1)$ and $b=(\eta_2,\phi_2)$, then the distance between these two points is
\begin{eqnarray} 
\Delta r_{ab} = \sqrt{(\eta_1-\eta_2)^2+(\phi_1-\phi_2)^2} \, .
\label{eq:Distance}
\end{eqnarray}
In the literature, we find several selection criteria for isolated photons, such as the cone isolation or smooth isolation prescription. Specifically for these two cases, studies have been carried out whose conclusions show the differences can be minimized or directly neglected for some observables \cite{Cieri:2015wwa,Catani:2018krb}. Due to its theoretical advantages, we used smooth isolation prescription ~\cite{Frixione:1998jh}. The selection procedure goes as follow:
\begin{enumerate}
    \item Identify each photonic signal in the final state, and draw a cone of radius $r_0$ around it.
    \item If there are no QCD partons inside the cone, the photon is isolated.
    \item If there are QCD partons inside the cone, we calculate their distance to the photon, $r_j$, following Eq. (\ref{eq:Distance}) and then we define the total transverse hadronic energy for a cone of radius $r$ as
    \begin{eqnarray}
    E_T(r) = \sum_{j} E_{T_j} \theta(r-r_j) \, ,
    \label{eq:Energy}
    \end{eqnarray}
    where $E_{T_j}$ is the transverse energy of the $j$-th QCD parton inside the cone.
    \item Define an arbitrary smooth function $\xi(r)$ which satisfies $\xi(r) \to 0$ for $r\to 0$.
    \item If $E_T(r)<\xi(r)$ for every $r<r_0$ (i.e. for any point inside the fixed cone), then the photon is isolated and the event is selected. Otherwise, we reject the event.
\end{enumerate}
This algorithm allows soft gluons in any region of the phase space, as well as the emission of soft-collinear massless quarks, leading to a IR-safe definition of the cross-section when higher order corrections are considered. In this way, the smooth isolation criteria allows to neglect the contribution due to resolved photons, described in Eq. (\ref{eq:Resolved}). Then,
\begin{eqnarray}
\nonumber d\sigma_{H_1 \, H_2 \to h \, \gamma} &=& \sum_{a_1 a_2 a_3} \int dx_1 dx_2 dz \, f^{(H_1)}_{a_1}(x_1,\mu_I) \\
&\times&f^{(H_2)}_{a_2}(x_2,\mu_I) \, d^{(h)}_{a_3}(z,\mu_F) \nonumber\\
&\times&d\hat\sigma^{\rm ISO}_{a_1\,a_2 \to a_3 \, \gamma}  \, ,
\label{eq:Isolated}
\end{eqnarray}
represents the full-contribution to the cross-section when only isolated prompt-photons are measured.

Once we have defined our observable, it is time to describe the corrections that we take into consideration for this work. Here, we are interested in the production of a direct photon plus a hadron. In particular, we use fragmentation functions to study the probability of producing a positively charged pion. If we consider the Born level kinematic process, the interaction is given by 
\begin{eqnarray}
H_1 (K_1) + H_2 (K_2) \to h (K_3) + \gamma (K_4) \, , 
\end{eqnarray}
with $K_i$ the four-momenta of the different particles in the lab frame. Let us first consider the cross-section calculation with NLO corrections from QCD. We relied on the FKS subtraction algorithm \cite{Frixione:1995ms} and the smooth isolation criteria in Eq. (\ref{eq:Isolated}), obtaining 
\begin{eqnarray}
 d\hat\sigma^{\rm ISO}_{a_1\,a_2 \to a_3 \, \gamma}&=&
  \frac{\alpha_s}{2\pi} \frac{\alpha}{2\pi}\, \int d{\rm PS}^{2\to 2} \,  \frac{|{\cal M}^{(0)}|^2}{2 \hat s} \, {\cal S}_2 \, 
\nonumber\\  &+& \frac{\alpha_s^2 }{4\pi^2} \frac{\alpha}{2\pi}\, \int d{\rm PS}^{2\to 2} \, \frac{|{\cal M}^{(1)}|^2}{2 \hat s} \,{\cal S}_2 \, 
\nonumber\\ &+& \frac{\alpha_s^2 }{4\pi^2} \frac{\alpha}{2\pi} \sum_{a_5} \int d{\rm PS}^{2\to 3} \, \frac{|{\cal M}^{(0)}|^2}{2 \hat s} \, {\cal S}_3 \,  ,
\nonumber \\ 
\,
\label{eq:xsISOLATEDQCD}
\end{eqnarray}
with $\hat s$ the partonic center-of-mass energy, $|{\cal M}^{(0)}|^2$ the squared matrix-element at Born level and $|{\cal M}^{(1)}|^2$ the corresponding one-loop contribution. ${\cal S}_2$ and ${\cal S}_3$ are the measure functions that implements the experimental cuts and the isolation prescription for the $2\to 2$ and $2 \to 3$ sub-processes, respectively. In addition, each of the amplitudes depends explicitly on the momentum of the partons and the particles in the final state, 
\begin{eqnarray}
 |{\cal M}^{(0)}|^2 &=& |{\cal M}^{(0)}|^2(x_1 K_1, x_2 K_2, K_3/z, K_4) \, ,\nonumber\\  
 |{\cal M}^{(1)}|^2 &=&|{\cal M}^{(1)}|^2(x_1 K_1, x_2 K_2, K_3/z, K_4) \, ,\nonumber\\
|{\cal M}^{(0)}|^2 &=&|{\cal M}^{(0)}|^2(x_1 K_1, x_2 K_2, K_3/z, K_4, k_5) \, ,
\label{eq:AMPLITUD}
\end{eqnarray}
where we introduce the momentum fractions $x_1$ and $x_2$ for the incoming partons, and $z$ for the final-state parton that hadronizes into $h=\pi^+$. For the LO contribution, there are two partonic channels,
\begin{equation}
q \bar q \to \gamma g \, , \quad q g \to \gamma q \, ,
\label{eq:PartonicChannelsLO}
\end{equation}
whilst, all the QCD channels contributing at NLO are given by,  
\begin{align}
q \bar q \to \gamma g g \, , \quad q g &\to \gamma g q \, , \quad g g  \to \gamma q \bar q \, , \nonumber\\
\quad q \bar q \to \gamma Q \bar Q \, &, \quad q Q \to \gamma q Q \, . 
\label{eq:PartonicChannelsNLO}
\end{align}

%20211230: REVISADO GS

%%%%%%%%%%%%%%%%%%%%%%%%%%%%%%%%%%%%%%%%%%%%%%%%%%%%%%%%%%%%%%%%%%%%%%%%%%%%%%%%%%%%%%%%%%%%%%%%%%%%%%%%
\section{Phenomenology and results}
\label{sec:results}
Using a Monte Carlo (MC) integrator, we implement Eq. (\ref{eq:xsISOLATEDQCD}) and simulate the production of a hadron plus a direct photon. The code as well as other technical details, were presented in Refs. \cite{deFlorian2011PRD83,RenteriaEstrada2021sym13}. For the smooth isolation algorithm, we use the function 
\begin{eqnarray}
\xi(r) = \epsilon_\gamma E_T^{\gamma} \, \left(\frac{1-\cos(r)}{1-\cos{r_0}}\right)^4 \, ,
\end{eqnarray} 
with the parameters $\epsilon_\gamma=1$, $r_0=0.4$ and the photon transverse energy $E_T^{\gamma}$. The average of the photon and hadron transverse energy was used as the typical energy scale of the process, i.e.
\begin{equation}
\mu \equiv \frac{p_T^{h}+p_T^{\gamma}}{2} \, ,
\end{equation}
and we set by default $\mu_I=\mu_F=\mu_R\equiv \mu$. Due to the fact that producing heavy hadrons is suppressed, we focus on the process,  
\begin{equation}
    p \, p \, \to \, \gamma \, + \pi^+ \, .
\end{equation}
The cuts used in the MC simulation are those established by the PHENIX detector from RHIC. Explicitly, 
\begin{itemize}
  \item Photon and pion rapidities: $\{|\eta^\gamma|,\,|\eta^\pi|\}\leq 0.35$.
  \item Photon transverse momentum: $5\, {\rm GeV} \leq p_T^{\gamma} \leq 15\, {\rm GeV}$.
  \item Pion transverse momentum: $p_T^{\pi} \geq\, 2 {\rm  GeV}$.
  \item Azimuth angle difference: $\Delta \phi=|\phi^\pi - \phi^\gamma|\geq 2$.
\end{itemize}
Regarding the center-of-mass energy of the hadron collisions, we use by default $E_{CM} = 200 \, {\rm GeV}$, although we also explored the TeV region accessible by LHC, setting $E_{CM} = 13 \, {\rm TeV}$. We are interested in studying the improvements associated to new and up-to-date PDF sets (for instance, the one developed by \texttt{NNPDF} collaboration \cite{AbdulKhalek:2019bux}) w.r.t. the predictions obtained with older distributions \cite{deFlorian:2009vb}. Likewise, we have suppressed, directly from the MC code, the interpolator of the distribution functions and we have carried out the interpolation through the LHAPDF framework\cite{Buckley:2014ana}. Besides, we implemented the updated set of FFs, \texttt{DSS2014} \cite{deFlorian:2014xna}. 

In order to study the partonic distribution effects, we consider three possible scenarios. First, we define our default configuration as $\sigma_a$, where we have implemented the most recent sets of PDFs. With this configuration, we use the \texttt{NNPDF3.1} PDFs \cite{Bertone:2017bme} and \texttt{DSS2014} fragmentations \cite {deFlorian:2014xna} from the \texttt{LHAPDF} framework. In the second scenario, i.e. $\sigma_b$, we explore the effects on the hadronization of the process. Thus, we take into account the modifications induced by changing the new FF to the old FF set from \texttt{DSS2007} \cite{deFlorian:2007ekg}, keeping the default choice for the PDFs. Finally, we define the $\sigma_c$ configuration, where we explore the effects in the parton distributions. Explicitly, we keep the default FF set, but switched to \texttt{MSTW2008NLO} PDFs \cite{Martin:2009iq}. 

The results obtained for RHIC at $E_{CM}=200\,{\rm GeV}$ are shown in Figs. \ref{fig:Figura1} and \ref{fig:Figura2}. In the left side of the plots, we present the differential cross-section as a function of $p_T^\pi$ and $p_T^{\gamma}$, exploring the range between 5 GeV and 15 GeV for both momenta. In the right side of the plots, we show the relative difference of $\sigma_b$ and $\sigma_c$ scenarios with respect to our default configuration. The analysis is performed taking into account the NLO QCD corrections. From the right plot in Fig. \ref{fig:Figura1}, we notice that the \texttt{MSTW2008} PDF tends to slightly enhance the high-$p_T$ region (i.e. $p_T^{\pi}\approx 13 \, {\rm GeV}$), whilst the effect of the \texttt{DSS2007} FF goes in the opposite direction. Namely, \texttt{DSS2007} gives a larger cross-section for low-$p_T$ and a smaller for high-$p_T$, reaching a relative difference of ${\cal O}(10 \, \%)$.

\myfigure{%
\includegraphics[width=1\columnwidth]{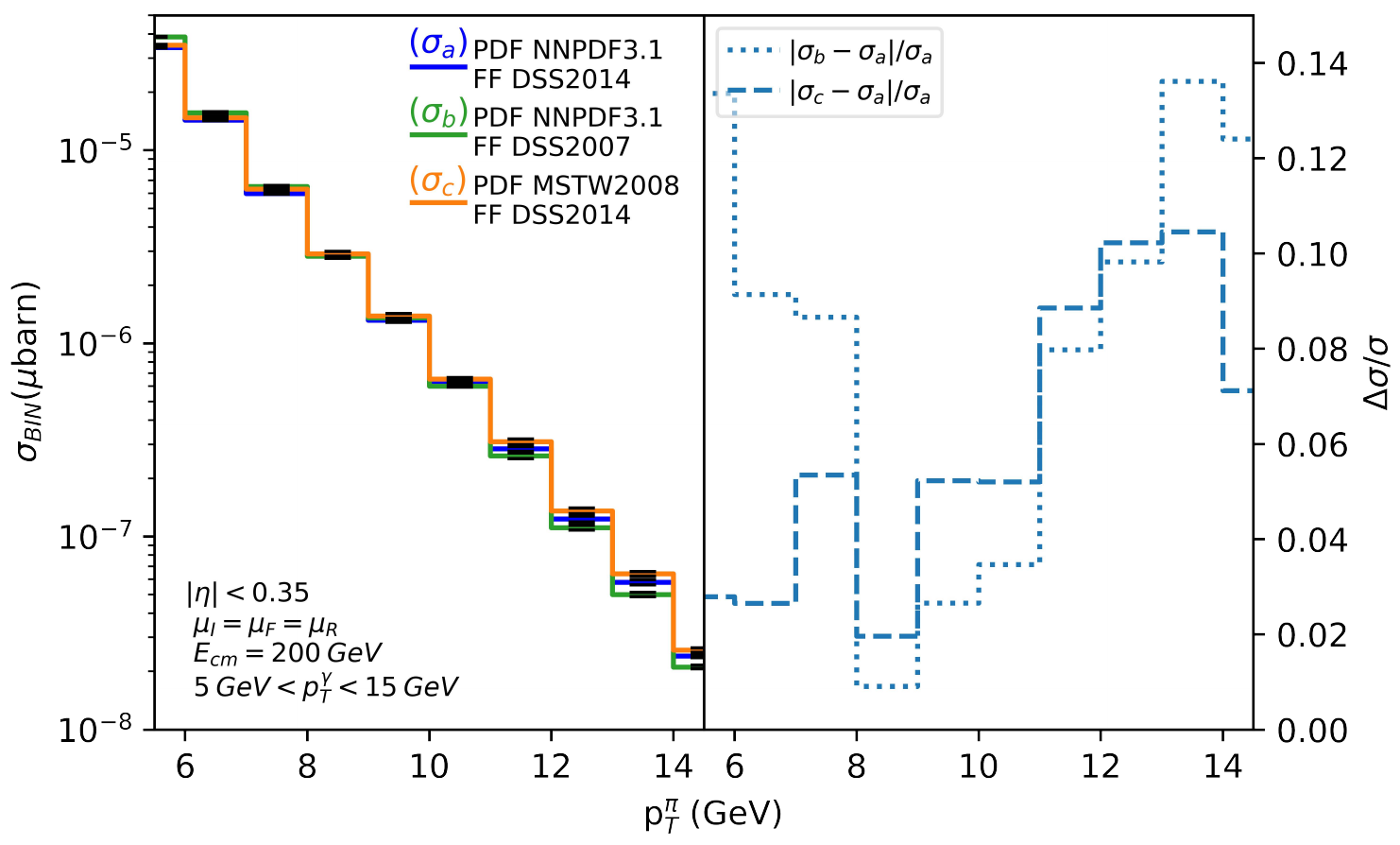}%
\figcaption{NLO corrections to the $p_T^{\pi}$ distribution for $pp\to \gamma + \pi^+$, for PHENIX kinematics ($E_{CM} = 200 \, {\rm GeV}$). In the right panel, we show the relative difference w.r.t. the default configuration.}%
\label{fig:Figura1}%
}
\myfigure{%
\includegraphics[width=1\columnwidth]{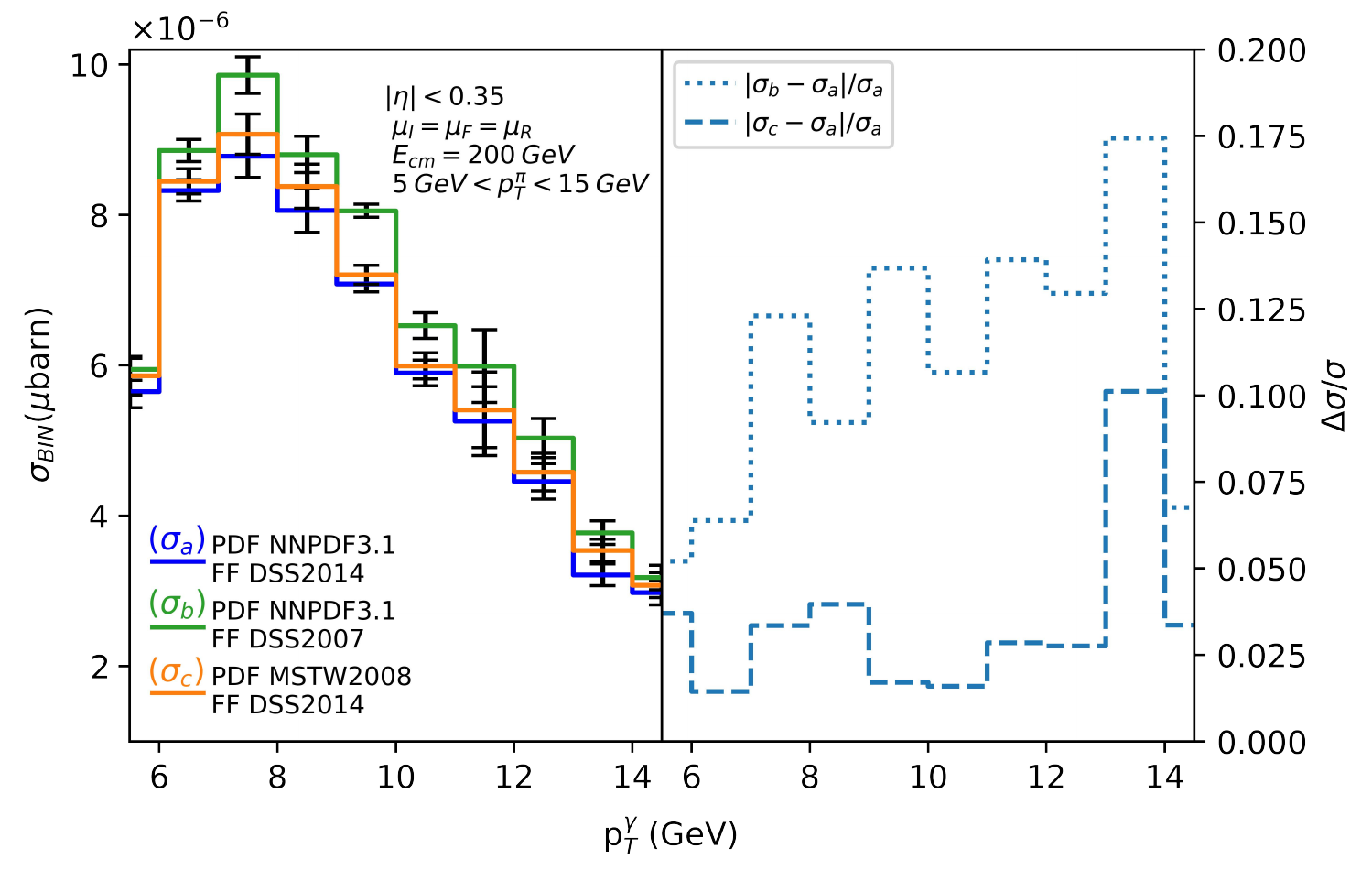}%
\figcaption{NLO corrections to the $p_T^{\gamma}$ distribution for $pp\to \gamma + \pi^+$, for PHENIX kinematics ($E_{CM} = 200 \, {\rm GeV}$). In the right panel, we show the relative difference w.r.t. the default configuration.}%
\label{fig:Figura2}%
}
On the other hand, we can look at Fig. \ref{fig:Figura2}, where we present the differential cross-section w.r.t. $p_T^{\gamma}$. Here, the right plot shows that the cross-section is about $12 \, \%$ bigger when using the \texttt{DSS2007} fragmentation, and the trend increases with the photon transverse momentum.  Finally, we also find that the cross-section is higher for the default configuration: the relative difference is ${\cal O}(7 \, \%)$.

For this work, we have also explored the cross-section distributions for LHC: the predictions are shown in Fig. \ref{fig:Figura3}. We have restricted the attention to $\sigma_a$ and $\sigma_c$. Namely, we explored the partonic distribution effects comparing \texttt{NNPDF3.1}  and \texttt{MSTW2008}, whilst keeping the same fragmentation functions. The differences are around $10 \, \%$, with an enhancement of the cross-section for $p_T^{\pi} \approx 13 \, {\rm GeV}$ for $\sigma_a$ w.r.t. $\sigma_c$. It is worth noticing that we did not use \texttt{DSS2007} for LHC energies because most of the events involve momentum fractions lying outside the validity range of the interpolator.

\myfigure{%
\includegraphics[width=1\columnwidth]{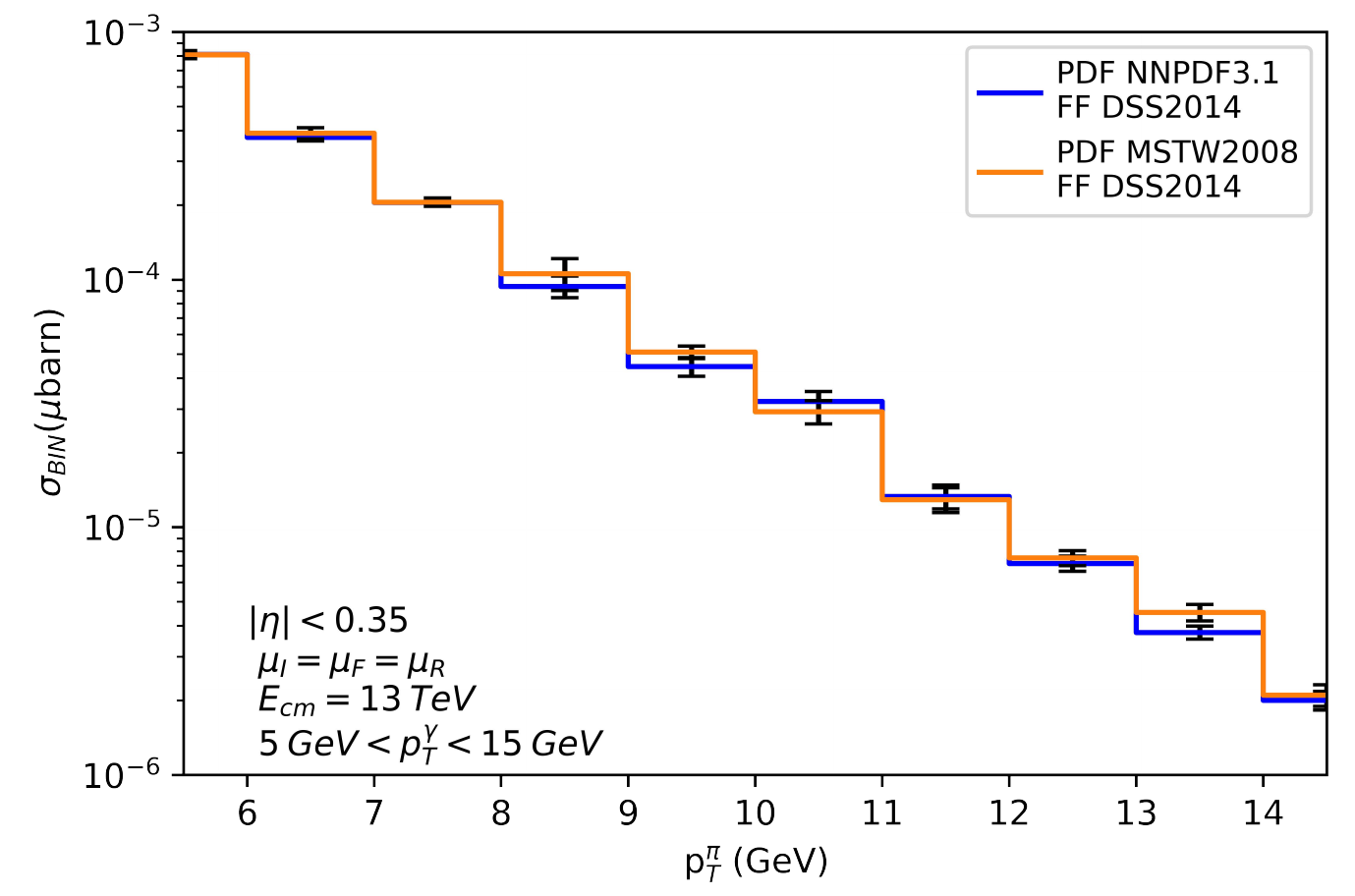}%
\figcaption{NLO  corrections to the $p_T^{\pi}$ distribution for $pp\to \gamma + \pi^+$, for LHC kinematics ($E_{CM} = 13 \, {\rm TeV}$, angular constraints identical to those used for PHENIX).}%
\label{fig:Figura3}}
%

%%%%%%%%%%%%%%%%%%%%%%%%%%%%%%%%%%%%%%%%%%%%%%%%%%%%%%%%%%%%%%%%%%%%%%%%%%%%%%%%%%%%%%%%%%%%%%%%%%%%%%%%%%%%
\section{Conclusion}
\label{sec:conclusions}
In this work, we have studied the phenomenology of the production of a direct photon plus a hadron, focusing on the process $pp\to\gamma+\pi^+$ calculated with up to NLO QCD accuracy. Our aim was to perform a comparison of the predictions when using the most recent set of partonic distribution functions w.r.t. the results presented in Ref. \cite{deFlorian2011PRD83}.

As a first step, we update our Monte Carlo by adapting the \texttt{LHAPDF} interface and updating the fragmentation to the new set \texttt{DSS2014}. We study the impact of changing the PDF and FF sets in the differential cross-section. In particular, we focused on the differences induced in the $p_T^{\pi}$ and $p_T^{\gamma}$ spectrum in the range of $5\,{\rm GeV} $ to $ 15\,{\rm GeV} $. We found reasonable deviations (i.e. ${\cal O} (10 \, \%)$ on average), although our preliminary studies suggest a stronger sensibility in the $p_T^{\gamma}$ distribution. 

The results presented in this article suggest that the production of a hadron in association with a hard photon could be useful for imposing more stringent restrictions on both PDF and FF. Likewise, the studies carried out in this work served as the basis Ref. \cite{RenteriaEstrada2021arxiv}, where advanced analysis were carried out with the purpose of reconstructing the internal kinematics of the parton-level collisions with the help of Machine Learning tools. 

%20211230: REVISADO GS

\section*{Acknowledgements}
This research was supported in part by COST Action CA16201 (PARTICLEFACE).
The work of D. F. R.-E. and R. J. H.-P. is supported by CONACyT (M\'exico) through the Project No. A1- S-33202 (Ciencia B\'asica) and Ciencia de Frontera 2021-2042; in addition by PROFAPI 2022 Grant No. PRO\_A1\_024 (Universidad Aut\'onoma de Sinaloa). Besides, R. J. H.-P. is also funded by Sistema Nacional de Investigadores from CONACyT.
\end{multicols}
\medline
\begin{multicols}{2}

\end{multicols}
\end{document}